\documentclass[final,3p,times,numbers,sort&compress]{elsarticle}   

\usepackage{amsmath,amssymb,stmaryrd,amstext,amsthm}
\usepackage{mathtools}
\usepackage{bm}	
\usepackage{caption}
\usepackage[hidelinks]{hyperref}
\usepackage{subcaption}
\usepackage{siunitx}
\usepackage{booktabs}
\usepackage[english]{cleveref}
\usepackage{lineno}
\usepackage{float}
\usepackage{graphicx}
\usepackage{verbatim}   
\newcommand{\volsym}{\rlap{\kern.08em--}V} 

\usepackage{verbatim}

\newcommand{\blue}{\color{black}} 

\newcommand{\black}{\color{black}}

\def\tsc#1{\csdef{#1}{\textsc{\lowercase{#1}}\xspace}}
\tsc{WGM}
\tsc{QE}
\tsc{EP}
\tsc{PMS}
\tsc{BEC}
\tsc{DE}

\journal{} 

\begin{document}

\begin{frontmatter}



\title{A progressive data-augmented RANS model for enhanced wind-farm simulations}
\author[label1]{Ali Amarloo}
\author[label1]{Navid Zehtabiyan-Rezaie}
\fntext[label1]{A. Amarloo and N. Zehtabiyan-Rezaie contributed equally to this study.}
\author{Mahdi Abkar\corref{cor1}}
\cortext[cor1]{Corresponding author}
\ead {abkar@mpe.au.dk}

\address{Department of Mechanical and Production Engineering, Aarhus University, 8200 Aarhus N, Denmark}

\begin{abstract}
The development of advanced simulation tools is essential, both presently and in the future, for improving wind-energy design strategies, paving the way for a complete transition to sustainable solutions. The Reynolds-averaged Navier-Stokes (RANS) models are pivotal in enhancing our comprehension of the complex flow within and around wind farms; hence, improving their capacity to accurately model turbulence within this context is a vital research goal. The enhancement is essential for a precise prediction of wake recovery and for capturing intricate flow phenomena such as secondary flows of Prandtl’s second kind behind the turbines. 
To reach these objectives, here, we propose a progressive data-augmentation approach. We first incorporate the turbine-induced forces in the turbulent kinetic energy equation of the widely used $k-\omega\text{SST}$ model. Afterward, we utilize data from large-eddy simulations to progressively enhance the Reynolds-stress prediction of this baseline model, accurately capturing the evolution of eddy viscosity in the wake, as well as the emergence of secondary flows. We then apply the optimized model to two unseen cases with distinct layouts and conduct a comparative analysis focusing on the obtained quantities such as normalized streamwise velocity deficit,  turbulence intensity, and power output. \blue We also evaluate the performance of the augmented model in predicting wake characteristics by comparing it with wind-tunnel measurement data\black. Our comparisons and validations demonstrate the superior performance of the progressive data-augmented model over the standard version in all cases considered in this study.
\end{abstract}



\end{frontmatter}

\section{Introduction} \label{Sec:Introduction}
The fluid mechanics of wind farms exhibit a significant degree of complexity associated with the high-dimensional nature of turbulence and the multi-scale interactions between the farm and the atmospheric boundary layer (ABL). Wakes -- the highly turbulent flow regions with reduced velocity forming behind turbines -- affect the lifetime and power production of the turbines operating downstream. Despite the turbulent-mixing mechanism diminishing wake effects over distance, the dense placement of turbines prevents full recovery of the flow features (see the review of Refs. \cite{Vermeer2003, Stevens2017, PorteAgel2019Review,ABKAR2023_TAML} and references therein). With the rising trends in wind energy, particularly the increase in power density of wind farms, the aforementioned challenge is set to become more pronounced. Given these considerations, acquiring in-depth knowledge and understanding of the wind-turbine and wind-farm wakes are 
key to advancing wind-energy deployment and facilitating society's full transition to green solutions \cite{Meneveau2019,Shapiro2021,papadis2020challenges}. 

The computational fluid dynamics (CFD) approach, via large-eddy simulation (LES) and Reynolds-averaged Navier-Stokes (RANS) models, has proven to be an invaluable tool for deepening our understanding of wind-farm-ABL interplay and, thus, facilitating a detailed exploration of the physics of flow inside and around wind farms (e.g., see reviews by Mehta \textit{et al.} \cite{Mehta2014} and Sanderse \textit{et al.} \cite{Sanderse2011}). While LES offers a high level of physical accuracy and aligns well with experimental and field data (see, e.g., Refs. \cite{PorteAgel2013, Wu2015,abkar2015influence, Stevens2018, AbkarEnergy2022}, among others), its computational demands make it better suited for addressing problems in fundamental research. In contrast, the RANS approach, being computationally more affordable, serves as the mainstream CFD tool in wind-energy applications, particularly in the industrial sector \cite{Eidi2021,TIAN2024}. 

The proficiency of RANS models in representing the intricate characteristics of wake flow hinges upon their turbulence-modeling accuracy, directly tied to their ability to predict the Reynolds-stress tensor (RST). Despite the popularity of RANS, it is well-established in the community that the common empirical models  (e.g., linear eddy-viscosity models like $k-\varepsilon$ and $k-\omega$ families) face critical challenges in the accurate prediction of eddy viscosity in turbine wakes which hampers their capabilities in delivering key quantities such as wake recovery and the output power of turbines \cite{Rethore2009,Eidi2022}. Motivated by this, empowering commonly used empirical models to accurately represent the eddy viscosity is a research objective pursued vigorously across numerous studies (see, e.g., Refs. \cite{ElKasmi2008, Prospathopoulos2011, VanDerLaan2015,Li2022,steiner2022classifying,Steiner2022,zehtabiyanrezaie2023extended}, among others). 
In addition to this, in our latest work \cite{zehtab2024secondary}, we highlighted that inaccurate prediction of eddy viscosity is not the only shortcoming of widely used RANS models, but it is also essential to enhance their capabilities to mimic the turbulence anisotropy. We showed the existence of secondary flows of Prandtl’s second kind \cite{nikitin2021prandtl} within turbine wakes using LES data. We attributed the upward-shifting trend in the wake center, evidenced by previous experimental and LES-based numerical observations \cite{chamorro2009wind,Chamorro2011,Abkar2016,zhang2023multiscale}, to the generation of secondary flows behind turbines. Our analysis revealed the critical role of spatial gradients of RST, particularly under the influence of ground effects, and tightly linked to the model's ability to predict turbulence anisotropy, in driving such flows. We demonstrated that linear-eddy viscosity models struggle to capture these complex flow features within wakes. 

While the limitations of RANS simulations in capturing the complexities of wind-farm flow physics remain an open challenge, encouragingly, the realm of wind energy is witnessing an increasing availability of data, rendering it a fertile ground for the application of data-driven and data-augmenting approaches for the development of next-generation models \cite{duraisamy2019turbulence, Brunton2022, zehtabiyan2022data}. 
Combining these endeavors with a progressive strategy has been demonstrated to establish models with high levels of accuracy, efficiency, and robustness  \cite{amarloo2023progressive,RINCON2023}. \blue A progressive approach introduces new corrections into conventional models to improve their performance in their specific well-established limitations, while preserving their original successful performance in other cases~\cite{xiang2022}. \black
Motivated by these advancements, here, we introduce a progressive data-augmented (PDA) RANS model tailored for the wake-flow simulation of wind farms. Our work aims to achieve two primary objectives: 1) improving the capabilities of a RANS model in predicting eddy viscosity and wake recovery behind the turbines, and 2) enabling it to capture turbulence anisotropy and mimic the formation of secondary flows of Prandtl’s second kind in the wake region. To this end, firstly, we incorporate the impact of turbine-induced forces in the turbulent kinetic energy (TKE) equation of the popular $k-\omega\text{SST}$ \cite{menter1994two} model, as proposed in Ref. \cite{zehtabiyanrezaie2023extended}. Building upon this baseline model, we leverage LES data and progressively refine the $\omega$-transport equation to better handle eddy-viscosity predictions \cite{amarloo2023progressive}, and introduce a non-linear term into its RST to mirror the emergence of secondary flows \cite{zehtab2024secondary}. 
\blue 
This study presents a comprehensive progressive approach to developing a modified RANS model, evaluating its potential for data-driven RANS modeling in addressing generalizability challenges. Additionally, it offers a practical and enhanced RANS model for simulating wind-farm flows.
\black This study encompasses the examination of three distinct wind-farm cases, with one serving as the basis for optimization and the remaining unseen cases utilized for model validation. We adopt the grid-search optimization technique to systematically explore a wide parameter space encompassing two key variables associated with the corrective terms. The performance of the new model is rigorously evaluated against LES data \blue and wind-tunnel measurements\black, serving as a benchmark for assessing its accuracy and robustness.

The rest of this paper is structured as follows: Section \ref{Sec:Method} introduces the methodology of the study, and in Section \ref{Sec:Results}, we discuss the results obtained from the baseline model and its PDA version. Finally, the key conclusions drawn from this study are summarized in Section \ref{Sec:Conclusions}.
\section{Methodology} \label{Sec:Method}
In this section, firstly, we introduce the baseline RANS model comprised of the $k-\omega\text{SST}$ \cite{menter1994two} with an extra term in its TKE equation accounting for the influence of turbine-induced forces \cite{zehtabiyanrezaie2023extended}. Afterward, the methodology adopted for progressive data augmentation of the baseline model is detailed. Subsequently, numerical setups of training and testing cases are introduced, followed by information on LESs.

\subsection{Progressive data-augmented RANS framework} 

By using the Reynolds decomposition of velocity and pressure, the RANS equations for an incompressible steady flow can be written as 

\begin{equation} \label{eq:Mass}
\partial_{i} \overline{u}_{i} = 0,
\end{equation}
\begin{equation} \label{eq:Momentum}
\overline{u}_{j} \partial_{j} \overline{u}_{i} = -\frac{1}{\rho } {\partial_{i} \overline{p}} + {\partial_{j}} \left( 2 \nu {S}_{ij} - {R}_{ij}\right)  + f_{i}, 
\end{equation}
where $i, j = 1, 2, 3$ indicate the streamwise ($x$), spanwise ($y$), and vertical ($z$) directions, respectively. $\overline{u}_{i}$ are the mean velocity components, $\overline{p}$ is the mean pressure, $S_{ij} = \frac{1}{2}(\partial_i \overline{u}_j  + \partial_j \overline{u}_i )$ is the mean strain-rate tensor, and ${R}_{ij}$ is the RST \cite{Pope2000}. Here, $f_{i}$ is turbine-induced forces determined by the actuator-disk model and defined as

\begin{equation} \label{eq:f_u}
f_{i} = -\frac{1}{2} C^\prime_T A_\text{cell} \left(\overline{u}_{\text{D},i} n_i\right)^2 \frac{\gamma_{j,k}}{V_\text{cell}}, 
\end{equation}

\noindent where $\overline{u}_{\text{D},i}$ is the disk-averaged velocity, and 
$C^\prime_T =4a /(1-a)$ 
denotes the disk-based thrust coefficient, in which $a$ represents the turbine's induction factor. Here, $n_i$ is the unit vector perpendicular to the disk, and $\gamma_{j,k}$ denotes the overlap fraction between the rotor and the cell at a grid point $(j,k)$. $A_\text{cell}$ and $V_\text{cell}$ are the frontal surface and volume of the computational mesh within the rotor placement, respectively \cite{calaf2010large}. 

We use the commonly used $k-\omega\text{SST}$ \cite{menter1994two} equipped with the effect of turbine-induced forces in its TKE equation \cite{zehtabiyanrezaie2023extended} as the baseline RANS model for the prediction of RST as  
\begin{equation} \label{eq:AijBL}
{R}^\text{BL}_{ij}=-2 \nu_\text{T} {S}_{ij} + \frac{2}{3} k \delta_{ij},
\end{equation}
\begin{equation}
\label{eq:nutModel}
\nu_t = \frac{a_1 k}{\text{max}(a_1 \omega, F_2 S)},
\end{equation}
where $k$ is the TKE, and $\omega$ is the specific dissipation rate. \blue The parameter $a_1$ is a model constant, with a value of 0.31, and $F_2$ is a blending function defined in the original $k-\omega\text{SST}$ model \cite{menter1994two}. \black In the baseline model, these quantities are calculated by two transport equations as
\begin{equation}
\label{eq:tkeModel}
    \partial_j \left (\overline{u}_j k \right ) = P_k - \beta^* \omega k + \partial_j \left[\left(\nu+\sigma_k \nu_t\right) \partial_j k \right] + S_k,
\end{equation}
\begin{equation}
\label{eq:omegaModel}
    \partial_j \left (\overline{u}_j \omega \right ) = \frac{\gamma}{\nu_t}P_k - \beta \omega^2  + \partial_j \left[\left(\nu+\sigma_{\omega}\nu_t\right)  \partial_j \omega \right]  +  CD_{k\omega}, 
\end{equation}
where $P_k = \partial_i\overline{u}_j (2\nu_t S_{ij})$ is the production of TKE by the RST. Compared to the standard version (for more details, see, e.g., Ref.~\cite{menter2003ten}), we introduce an extra term, denoted as $S_k$, that corresponds to the impact of turbine-induced forces on the production of TKE and is calculated as

\begin{equation} \label{eq:Sk}
S_k \approx -\frac{1}{2} C^\prime_T A_\text{cell} \left[\frac{4}{3} k_\text{D} \overline{u}_{\text{D},x} + \left(\frac{2}{3} k_\text{D}\right)^{3/2} \right]\frac{\gamma_{j,k}}{V_\text{cell}}, 
\end{equation}

\noindent where $k_\text{D}$ is the disk-averaged TKE. 
The interested reader is referred to Ref.~\cite{zehtabiyanrezaie2023extended} for further details on this modification of the TKE equation.

Building upon our prior study \cite{zehtab2024secondary} and the progressive augmentation of RANS models in Refs.~\cite{RINCON2023, amarloo2023progressive}, a correction term is considered in Eq. (\ref{eq:AijBL}) for capturing the Prandtl's secondary flows of the second kind as

\begin{equation} \label{eq:AijSF}
{R}_{ij}=-2 \nu_\text{T} \left[{S}_{ij} - C_\text{SF} \frac{\left( {S}_{ik}{\Omega}_{kj}-{\Omega}_{ik}{S}_{kj}\right)}{\omega}\right] + \frac{2}{3} k \delta_{ij}, 
\end{equation}
where $\Omega_{ij} = \frac{1}{2}(\partial_i \overline{u}_j  - \partial_j \overline{u}_i )$ is the mean rotation-rate tensor, and $C_{\text{SF}}$ is the secondary-flow coefficient which will be optimized in this study. 

Inspired by the progressive augmentation of $k-\omega\text{SST}$ in Ref. \cite{amarloo2023progressive}, we also consider a correction term for the production source of $\omega$-transport equation which will augment the RANS model to deliver more accurate eddy viscosity in the near- and far-wake regions. Thus, Eq. (\ref{eq:omegaModel}) is modified as 

\begin{subequations}
\label{eq:PDAomega}
    \begin{equation}
    \label{eq:progressiveOmegaModel}
        \partial_j \left (\overline{u}_j \omega \right ) = \frac{\gamma}{\nu_t}P_k \left (1 + C_{P\omega}\mathcal{X} \right ) - \beta \omega^2  + \partial_j \left[\left(\nu+\sigma_{\omega}\nu_t\right) \partial_j \omega \right] +  CD_{k\omega},
    \end{equation}
    \begin{equation}
    \label{eq:separationFactor}
        \mathcal{X} = 1 - \nu_t \frac{\omega}{k},
    \end{equation}
\end{subequations}
where $C_{P\omega}$ is the strength of the correction for the production term in the $\omega$-transport equation which will be optimized in this study. Here, $\mathcal{X}$ is a linear activation function based on shear stress transport (SST) reflection inside the $k-\omega\text{SST}$ model. \blue 
It should be noted that $\mathcal{X}$ is only activated in regions where the ratio of production to dissipation of $k$ significantly deviates from one and, consequently, the definition of $\nu_t = {k}/{\omega}$ is no longer valid \cite{menter1994two}.
\black 
The reader may refer to Ref.~\cite{amarloo2023progressive} for further details on this modification. Hereafter, we will refer to our proposed model as $k-\omega\text{SST}-\text{PDA}$.

\subsection{Description and simulation setup of wind-farm cases}
\label{sec:numSetup}
\begin{figure}[!ht]
	\centering
        \includegraphics[width=0.85\textwidth]{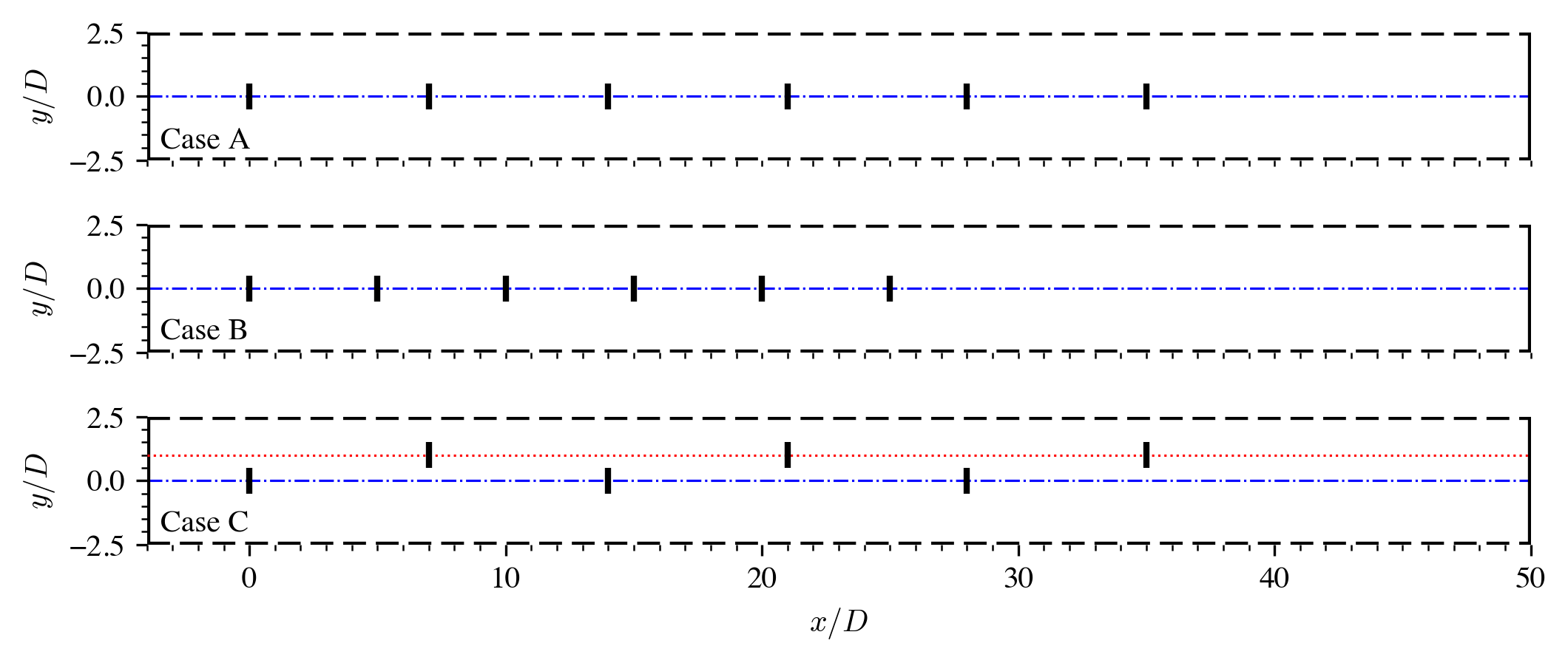}
	\caption{Schematics of three wind-farm layouts considered in this study. The turbines' positions are indicated by black rectangles, while dashed lines denote the cyclic boundaries.}
	\label{fig:Layout}
\end{figure}

In this study, we consider three different wind-farm layouts (Cases A, B, and C) where we use Case A for the training (i.e., the optimization of $C_{\text{SF}}$ and $C_{P\omega}$ values in $k-\omega\text{SST}-\text{PDA}$), and Cases B and C are considered for the validation of the optimized model. Figure \ref{fig:Layout} shows the layout of cases with 6 rows of turbines. In all cases. the first turbine row is located at the distance of $5D$ from the inlet, where $D = 80 \text{ m}$ is the rotor diameter. The streamwise spacing is $7D$ for Cases A and C while being $5D$ for Case B. In Case C, the even turbine rows are shifted in the spanwise direction by $1D$. The turbines in all cases operate with an induction factor of $a = 0.25$. The inflow condition has a velocity of $8$ m/s and turbulence intensity of $5.8$\% at the hub height of $z_\text{hub} = 70$ m. 

For the RANS simulations, we use the \texttt{simpleFoam} solver within the open-source OpenFOAM-v2312 package \cite{weller1998tensorial}. We implement the new turbulence model to solve the governing equations introduced earlier, enabling us to simulate turbine wakes and power output under a neutrally stratified ABL. The computational domain has dimensions of $4400\text{ m} \times 400\text{ m} \times 355\text{ m}$ in streamwise, spanwise, and vertical directions, respectively. The resolution of $234\times40\times58$ is adequate according to \blue our grid-convergence study (see \ref{app:App})\black. The top boundary has symmetry boundary conditions, cyclic conditions are used for side boundaries, and a rough-wall boundary condition with wall models is used for the bottom wall. \blue The reader may refer to Ref.~\cite{yang2009new} for more information about this wall model\black. The RANS inflow is aligned with that of LES, and more details on the procedure can be found in Ref. \cite{zehtab2024secondary}.

The high-fidelity data of these cases is needed in both the training and testing phases. For this purpose, we use an in-house pseudo-spectral finite-difference code for the LES of the three cases. This LES framework has been frequently used and validated (see, e.g., Refs. \cite{Abkar2014,abkar2015influence,abkar2015new,Abkar2016}), and for the sake of brevity, we skip further details on the LES numerical setup. Interested readers may refer to Ref. \cite{Eidi2022} for more information.

\section{Results and discussion} \label{Sec:Results}
In this section, initially, we present the simulation-driven optimization process conducted to derive the optimal corrective coefficients within the core of the $k-\omega\text{SST}-\text{PDA}$ model (i.e., $C_{\text{SF}}$ and $C_{P\omega}$). We first apply the model to Case A to evaluate its performance in the training case. Afterward, the performance of the novel progressive model is validated against LES when applied to unseen wind-farm cases. \blue Lastly, we focus on comparing the evolution of streamwise velocity downstream of several rows in a $10 \times 3$ turbine array. This comparison is benchmarked against wind-tunnel data from Chamorro and Port{\'e}-Agel \cite{Chamorro2011} and LES results from Stevens \textit{et al.} \cite{Stevens2018}\black.

\subsection{Simulation-driven optimization}
As given in Eq. (\ref{eq:AijSF}) and (\ref{eq:progressiveOmegaModel}), there are two coefficients including $C_{\text{SF}}$ and $C_{P\omega}$ which are yet to be determined to construct $k-\omega\text{SST}-\text{PDA}$. \blue For this purpose, we systematically search in a range of coefficients between $0$ and $2$ and evaluate the performance of $k-\omega\text{SST}-\text{PDA}$ on Case A. We have used a grid-search optimization where both coefficients with increments of $0.05$ are evaluated, demanding $41\times 41$ simulations. \black To quantify the success rate, we define a volumetric-averaged error as
\begin{equation}
    e_{\phi} = \frac{\int_{}^{} \int_{}^{} \int_{}^{} | \overline{\phi} - \overline{\phi}^{\text{LES}} | \text{ d}x\text{d}y\text{d}z}{ \int_{}^{} \int_{}^{} \int_{}^{} | \overline{\phi}^{\text{LES}} | \text{ d}x\text{d}y\text{d}z}\times 100,
    \label{eq:VolumetricError}
\end{equation}
where $\phi$ is a field quantity of interest and, in this study, we consider features like velocity components and turbulence intensity. 
The integration is performed in a finite box around the column of turbines where an accurate prediction of quantities of interest is essential. We also calculate the error in the prediction of power output via

\begin{equation}
    e_{P} = \frac{\Sigma|{P}_i - {P}^\text{LES}_i|}{\Sigma{P}^\text{LES}_i}\times 100,
    \label{eq:powerError}
\end{equation}
where ${P}_i$ indicates the power output corresponding to the $i$th turbine row.

\begin{figure}[!ht]
	\centering
        \includegraphics[width=\textwidth]{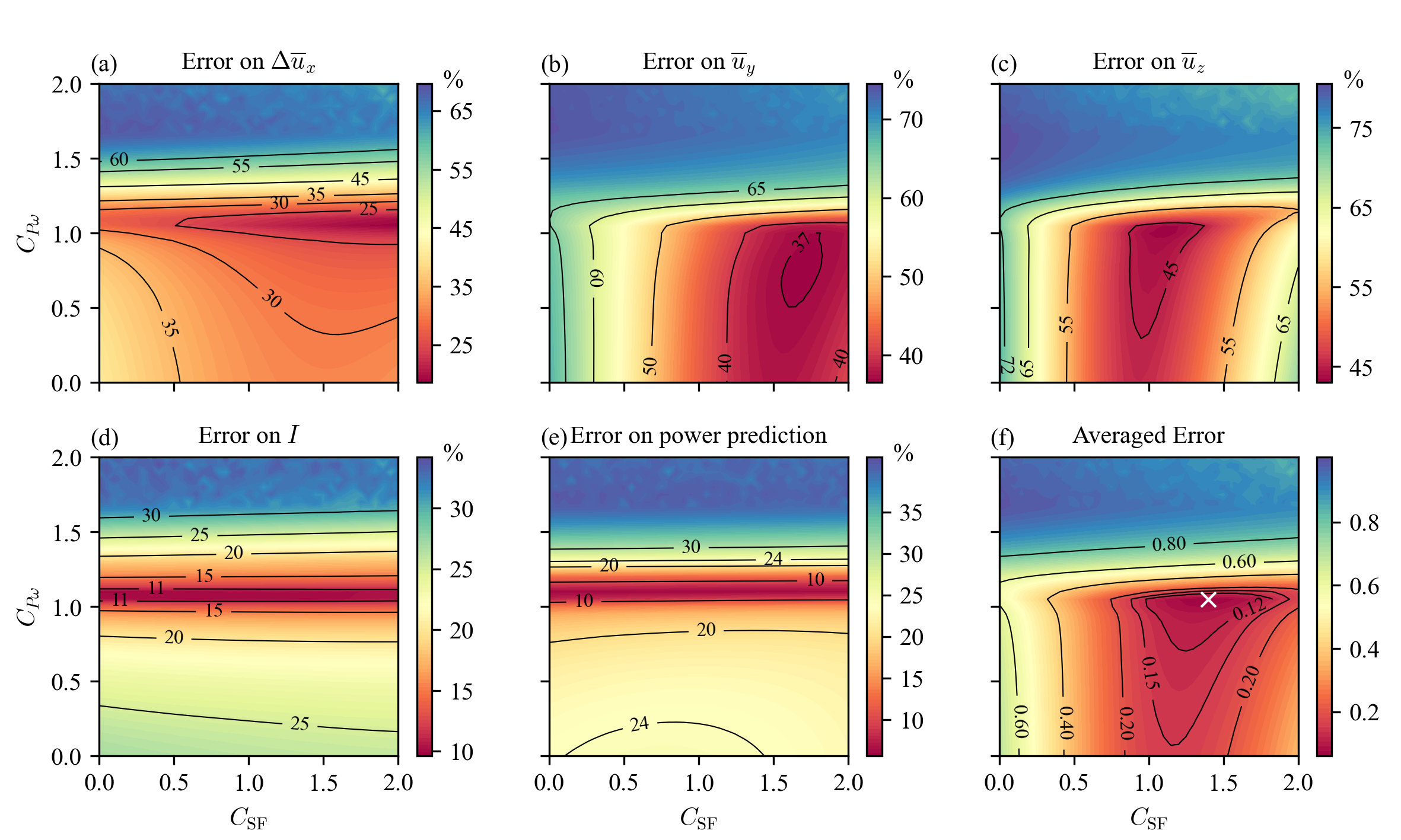}
	\caption{\blue Volumetric-averaged error values on (a) streamwise velocity deficit, (b) spanwise velocity, (c) vertical velocity, and (d) turbulence intensity. (e) The error in prediction of power output. Contour (f) shows the overall averaged error on velocity prediction (defined in Eq. (\ref{eq:avgError})). The white cross indicates the selected coefficients for the optimized model corresponding to the minimum value of the overall averaged error.\black}
	\label{fig:optimization}
\end{figure}

Figure~\ref{fig:optimization} presents the performance of $k-\omega\text{SST}-\text{PDA}$ as a function of different values of $C_{\text{SF}}$ and $C_{P\omega}$. Figures~\ref{fig:optimization}(a-c) show the volumetric-averaged errors on velocity components including $\Delta \overline{u}_x = (\overline{u}_{x,\text{in}} - \overline{u}_x)$, $\overline{u}_y$, $\overline{u}_z$, respectively, with $\overline{u}_{x,\text{in}}$ being the inflow velocity.
Figure~\ref{fig:optimization}(a) indicates that incorporating $C_{P\omega}$ into the $\omega$-transport equation significantly influences the prediction of the streamwise velocity (tightly linked to the model's success in eddy-viscosity predictions) while the impact of $C_{\text{SF}}$ is minimal. On the other hand, Figures~\ref{fig:optimization}(b) and (c) show that integrating the nonlinear term into RST will augment the baseline model with the prediction of secondary flows; hence, a noteworthy reduction of error in the prediction of $\overline{u}_y$ and $\overline{u}_z$ can be achieved by selecting an appropriate value for $C_{\text{SF}}$. Figure~\ref{fig:optimization}(d) presents the volumetric-averaged error in the prediction of turbulence intensity ($I = \sqrt{2k/3}/U_\text{hub}$, with $U_\text{hub}$ as the hub-height inflow velocity) which indicates that a better prediction of turbulence intensity is under the effect of $C_{P\omega}$ rather than $C_{\text{SF}}$. Similar behavior is seen for the error in the prediction of turbines' power output, as shown in Figure~\ref{fig:optimization}(e).

To determine the optimal combination of $C_{\text{SF}}$ and $C_{P\omega}$ for achieving the most accurate reconstruction of all velocity components, we establish an overall averaged error by considering the errors associated with the velocity components as

\begin{equation}
    e = \frac{1}{3}\Sigma_{n=1}^{3} \frac{e_n - \text{min}(e_n)}{\text{max}(e_n) - \text{min}(e_n)},
    \label{eq:avgError}
\end{equation}
where $e_n$ is the volumetric-averaged error on $\Delta \overline{u}_x$, $\overline{u}_y$, and $\overline{u}_z$. Figure~\ref{fig:optimization}(f) shows the contour of the overall averaged error ($e$), where the point with minimum value (indicated with the white cross) is selected as the optimized model coefficients with $C_{\text{SF}} = 1.4$ and $C_{P\omega} = 1.05$. 

\begin{figure}[!ht]
	\centering
        \includegraphics[width=\textwidth]{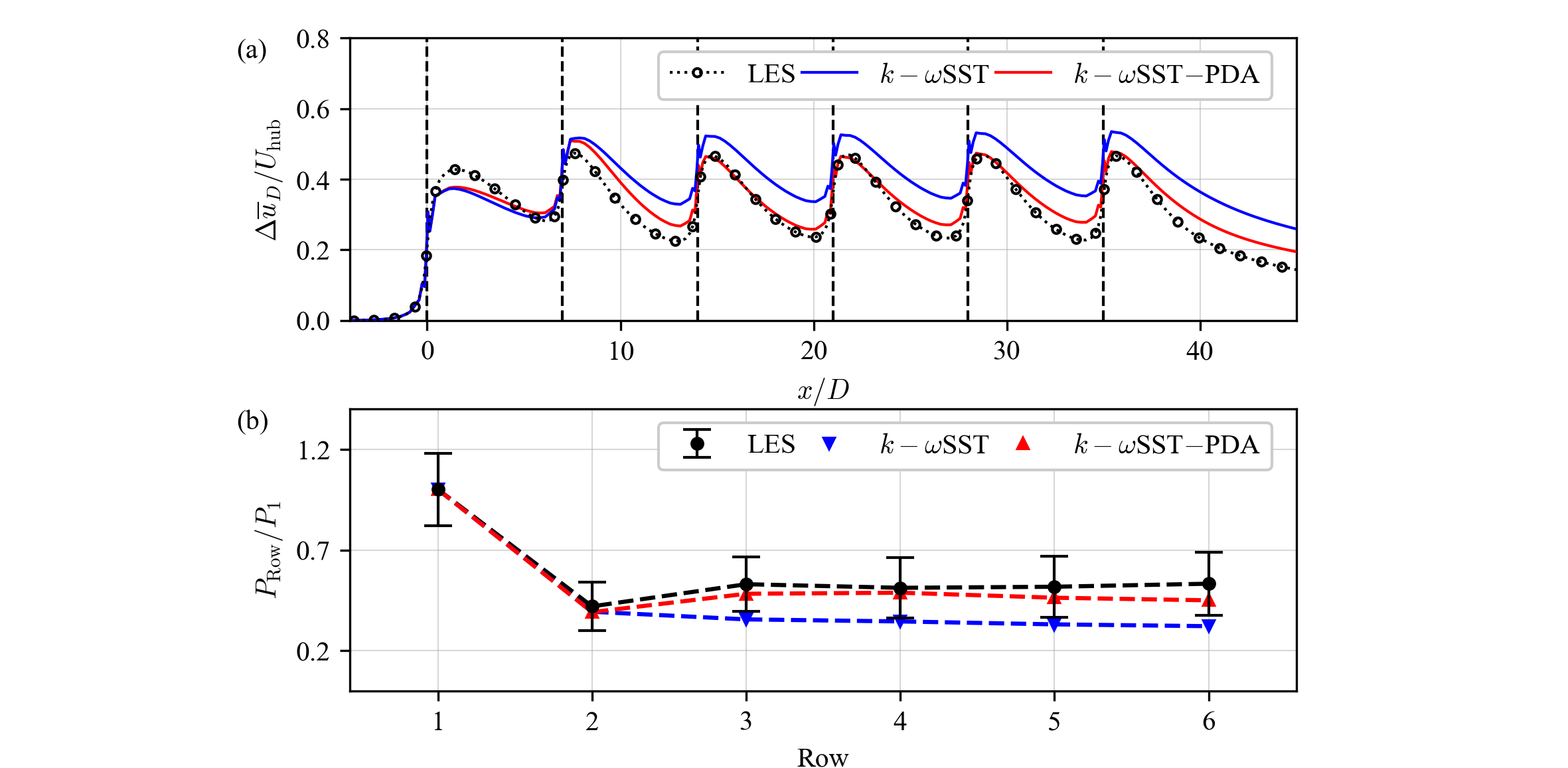}
	\caption{(a) Normalized rotor-averaged velocity deficit where the vertical dashed lines indicate the position of turbines, and \blue (b) the normalized power for each turbine row, both given for Case A\black. The normalized power is calculated by dividing the output power of each row by that of the first row. The vertical bars in (b) indicate the standard deviations of LES results.}
	\label{fig:WFA_DiskQuantities}
\end{figure}

For a better investigation of the performance of the $k-\omega\text{SST}-\text{PDA}$ with optimal coefficients, firstly, we apply it to the training case itself. Figure~\ref{fig:WFA_DiskQuantities}(a) shows the rotor-averaged streamwise velocity deficit, where rotor-averaging is conducted across the shared region of the $x$-normal planes and a hypothetical cylinder originating from the inlet, possessing the turbines' diameter and aligned along the hub height axis. The results show that $k-\omega\text{SST}$ overpredicts the velocity deficit, and this deficiency is addressed by the $k-\omega\text{SST}-\text{PDA}$ model. As highlighted in our recent study \cite{zehtab2024secondary} and also verified in Figures~\ref{fig:optimization}(a), (d), and (e), capturing secondary flows does not significantly alter the wake recovery and, thus, this enhancement is mainly attributed to the correction applied to the transport equation of $\omega$ which is also verified with the findings in Figure~\ref{fig:optimization}. 
\blue Figure~\ref{fig:WFA_DiskQuantities}(a) also illustrates that up to $x/D = 0$, where the first turbine is located, the $k-\omega\text{SST}-\text{PDA}$ model behaves identically to the original $k-\omega\text{SST}$ model, as expected, since the activation function is not yet enabled. Beyond this point, the activation function introduces the correction term, leading to noticeable discrepancies between the $k-\omega\text{SST}$  and $k-\omega\text{SST}-\text{PDA}$  models. This effect accumulates further downstream, becoming more pronounced after $x/D \simeq 10$\black

Focusing on the output-power predictions, shown in Figure~\ref{fig:WFA_DiskQuantities}(b), while the baseline model successfully captures the normalized power of the second row, it consistently underestimates for the turbine rows 3 to 6, linked to under-prediction of wake recovery by this model deep inside the farm. The data augmentation successfully improves this shortcoming of the baseline model while maintaining its satisfactory performance on the second row as a consequence of our progressive approach.

\begin{figure}[!ht]
	\centering
        \includegraphics[width=\textwidth]{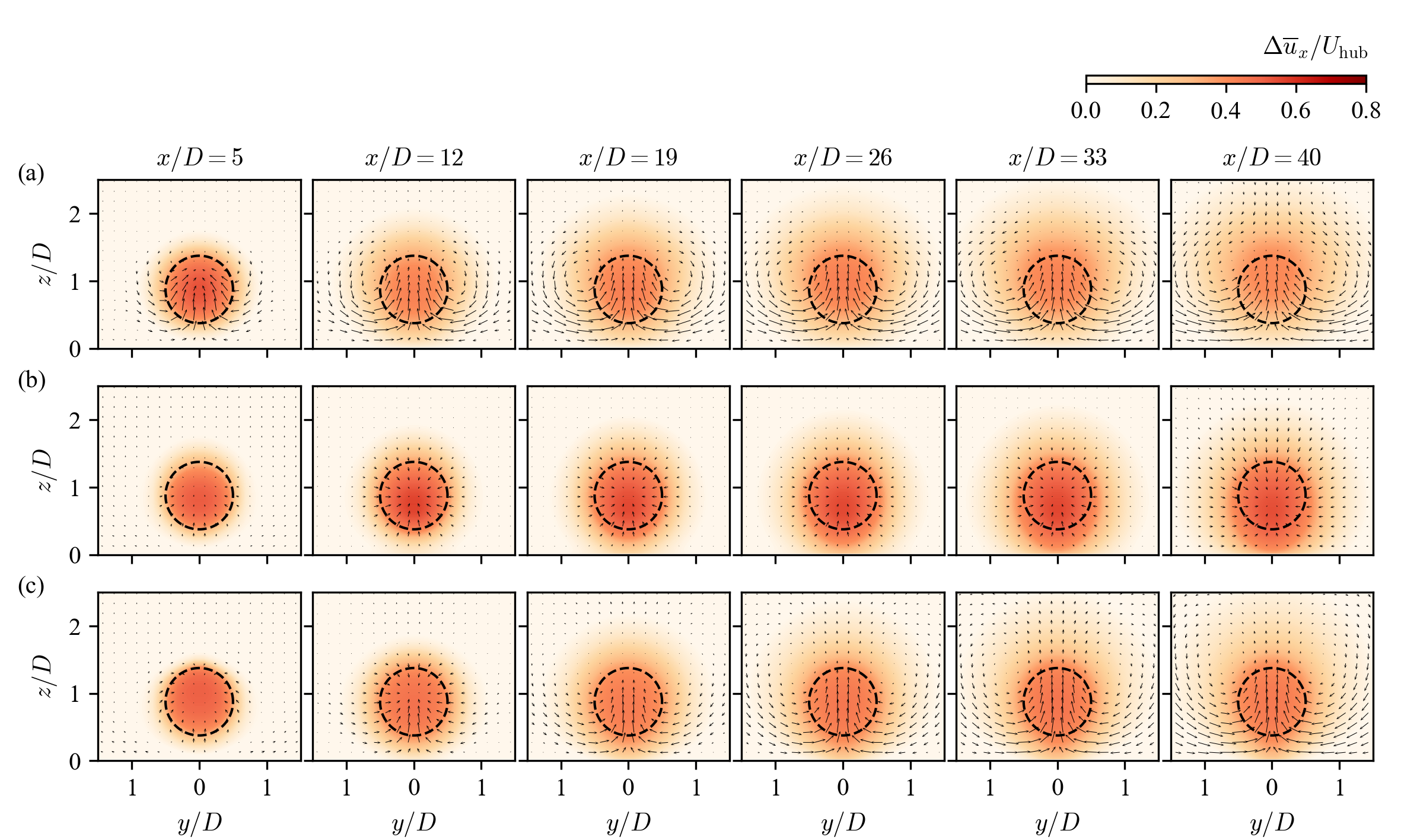}
	\caption{Contours of normalized streamwise velocity deficit and arrows of in-plane velocity, $5D$ after each turbine in Case A obtained from (a) LES, (b) $k-\omega\text{SST}$, and (c) $k-\omega\text{SST}-\text{PDA}$. Here, the black dashed circles indicate the rotor.}
	\label{fig:WFA_SFplots}
\end{figure}

Figure~\ref{fig:WFA_SFplots} compares the in-plane motion and the normalized streamwise velocity deficit between LES, $k-\omega\text{SST}$, and $k-\omega\text{SST}-\text{PDA}$, at a distance of $5D$ downstream of turbines in Case A. The main observation in this comparison is that the incorporation of the non-linear term into the RST results in a successful reconstruction of the secondary flows by using the $k-\omega\text{SST}-\text{PDA}$ with $C_{\text{SF}} = 1.4$, especially from the third row onward. Capturing such flow structures enables a shift of wake center towards a behavior that closely resembles what is observed in LES (Figure~\ref{fig:WFA_SFplots}(a)). Also, comparing Figures~\ref{fig:WFA_SFplots}(b) and (c) shows that the correction of the $\omega$-transport equation with $C_{P\omega} = 1.05$ enhances the values predicted for the normalized velocity deficit, compensating for under-prediction of wake recovery inside the farm by the baseline model.

\begin{figure}[!ht]
	\centering
        \includegraphics[width=0.833\textwidth]{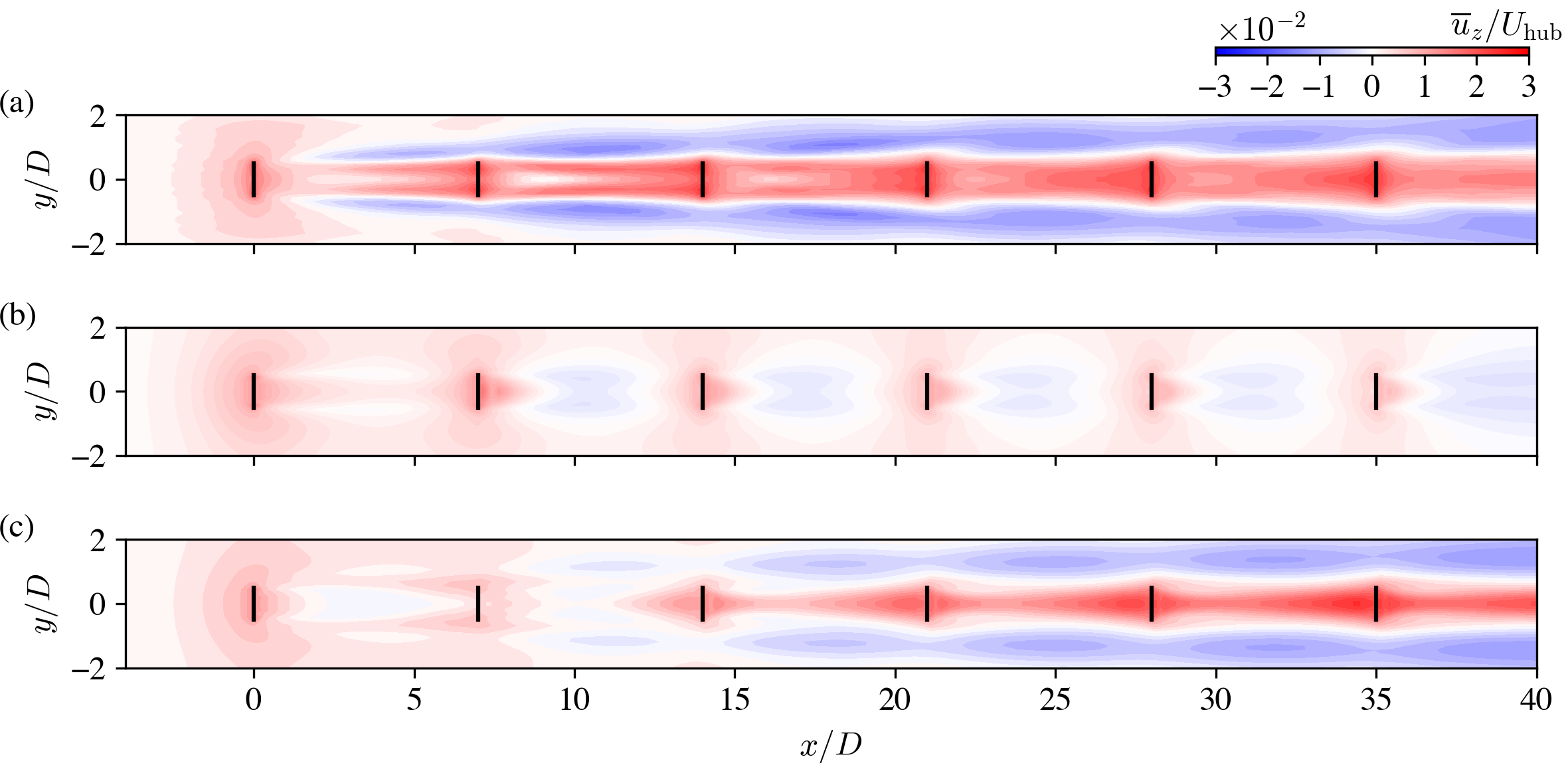}
	\caption{Contours of normalized vertical velocity at hub height in Case A obtained from (a) LES, (b) $k-\omega\text{SST}$, and (c) $k-\omega\text{SST}-\text{PDA}$.}
	\label{fig:WFA_uzHubHeight}
\end{figure}

For better observation of the effect of the RST's non-linear term with $C_{\text{SF}} = 1.4$ on the prediction of the secondary flows, we investigate the normalized vertical velocity component in a $xy$-plane at hub height, as given in Figure~\ref{fig:WFA_uzHubHeight}. The LES result indicates the dominance of a downward flow on the sides and an upward flow in the core of the wake, which is a consequence of the existence of the secondary flows. While the $k-\omega\text{SST}-\text{PDA}$ model satisfactorily mimics a closer-to-LES behavior, especially after the third turbine row, such flow characteristics are not observed in predictions made by the baseline model. We can reasonably extend the findings in this figure to Cases B and C, and for conciseness, we omit the results for the normalized vertical velocity component in those cases.

\begin{figure}[!ht]
	\centering
        \includegraphics[width=0.833\textwidth]{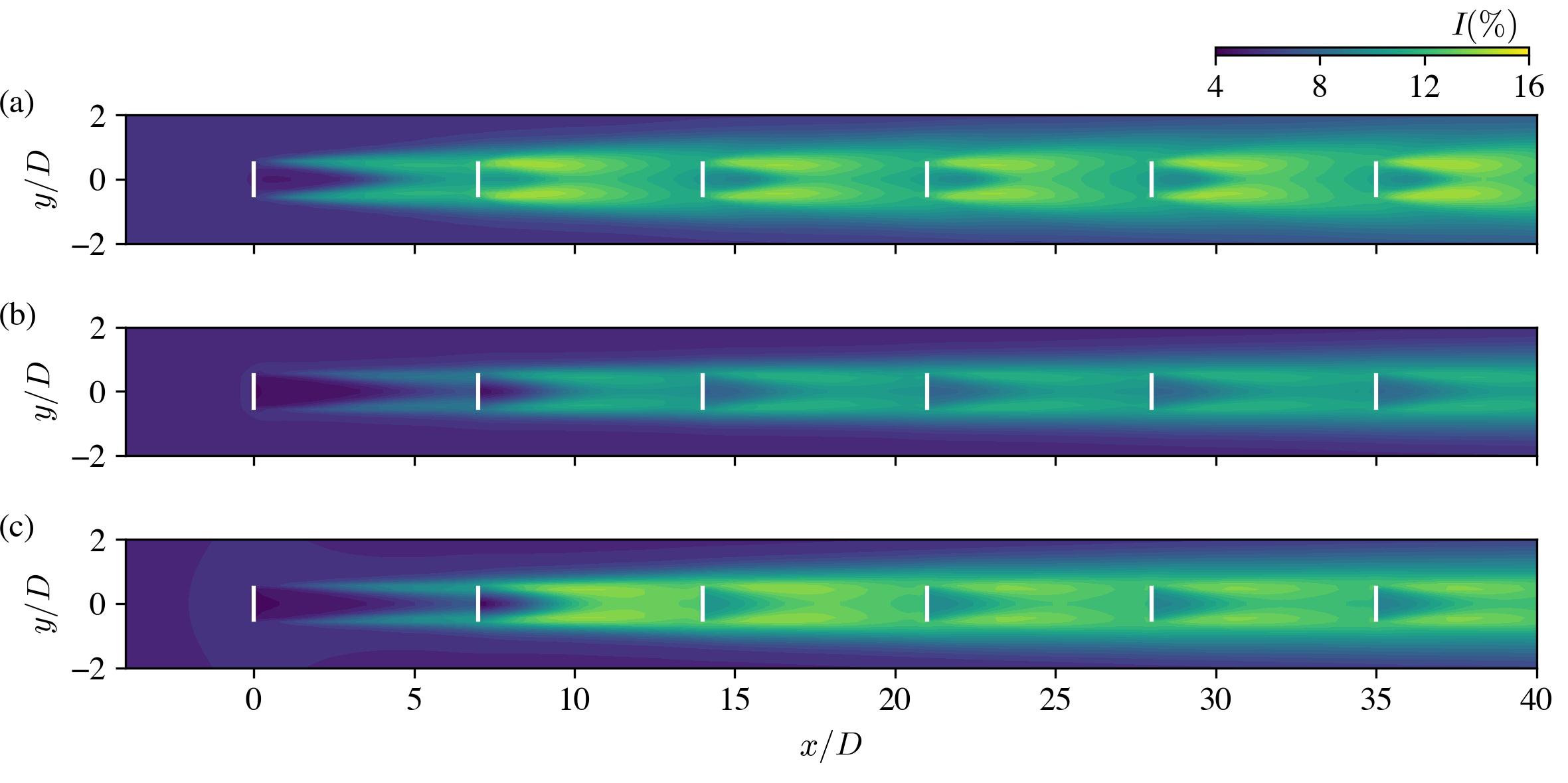}
	\caption{Contours of turbulence intensity at hub height in Case A obtained from (a) LES, (b) $k-\omega\text{SST}$, and (c) $k-\omega\text{SST}-\text{PDA}$.}
	\label{fig:WFA_Intensity}
\end{figure}

Figure~\ref{fig:WFA_Intensity} shows the distribution of turbulence intensity in a $xy$-plane at hub height, obtained from LES, baseline model, and $k-\omega\text{SST}-\text{PDA}$. While the baseline model can capture the double-peak structure of turbulence intensity behind the turbines, it underpredicts the magnitude of turbulence intensity, resulting in a weak recovery of wake. This justifies the overpredicted velocity deficit obtained from this model as observed in Figure~\ref{fig:WFA_DiskQuantities}(a). On the contrary, using $k-\omega\text{SST}-\text{PDA}$  with $C_{\text{SF}} = 1.4$ and $C_{P\omega} = 1.05$ delivers a closer-to-LES magnitude of turbulence intensity. 

So far, we can conclude that our approach yields multifaceted success; not only in capturing secondary flows and more accurately obtaining velocity deficit but also in accurately predicting the other important wake-flow feature, i.e., turbulence intensity. 

\subsection{Validation on testing cases} 

To assess the $k-\omega\text{SST}-\text{PDA}$ model's accuracy and robustness in unseen scenarios, in this section, we apply it to Cases B and C. It should be mentioned that in these validation cases, the same optimized values of $C_{\text{SF}} = 1.4$ and $C_{P\omega} = 1.05$ are fed into the PDA model to measure the level of generalizability and robustness achieved by the progressive data-augmentation approach. It is worth reminding that in Case B, the distance between turbine rows is reduced compared to Case A; thus, turbines experience a more intense wake effect. In Case C, the even turbine rows are shifted $1D$ in the spanwise direction causing the downstream turbines to experience partial wake.

\begin{figure}[!ht]
	\centering
        \includegraphics[width=\textwidth]{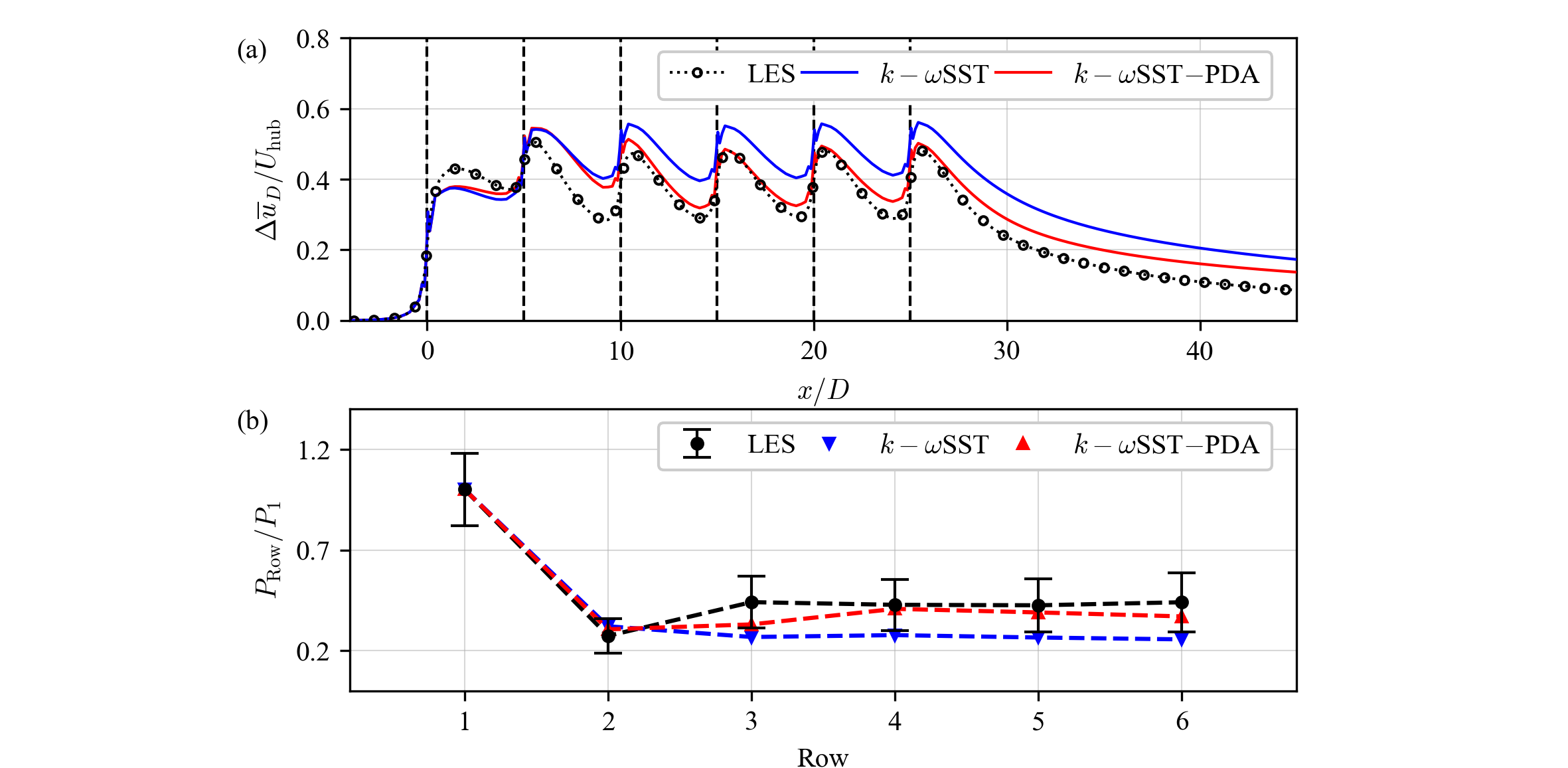}
	\caption{(a) Normalized rotor-averaged velocity deficit, and \blue (b) the normalized power for each turbine row, both given for Case B.\black}
	\label{fig:WFB_DiskQuantities}
\end{figure}

Concentrating on the rotor-averaged streamwise velocity deficit in Case B (Figure~\ref{fig:WFB_DiskQuantities}(a)), one can observe an overestimation of velocity deficit, showing that here, similar to Case A, the baseline model suffers from the underestimation of wake recovery. Turning to the $k-\omega\text{SST}-\text{PDA}$ model, the corrected $\omega$-transport equation improves the prediction of the velocity deficit, which leads to a satisfactory agreement in the prediction of power outputs, aligning well with the results obtained from LES (Figure~\ref{fig:WFB_DiskQuantities}(b)). 

\begin{figure}[!ht]
	\centering
        \includegraphics[width=\textwidth]{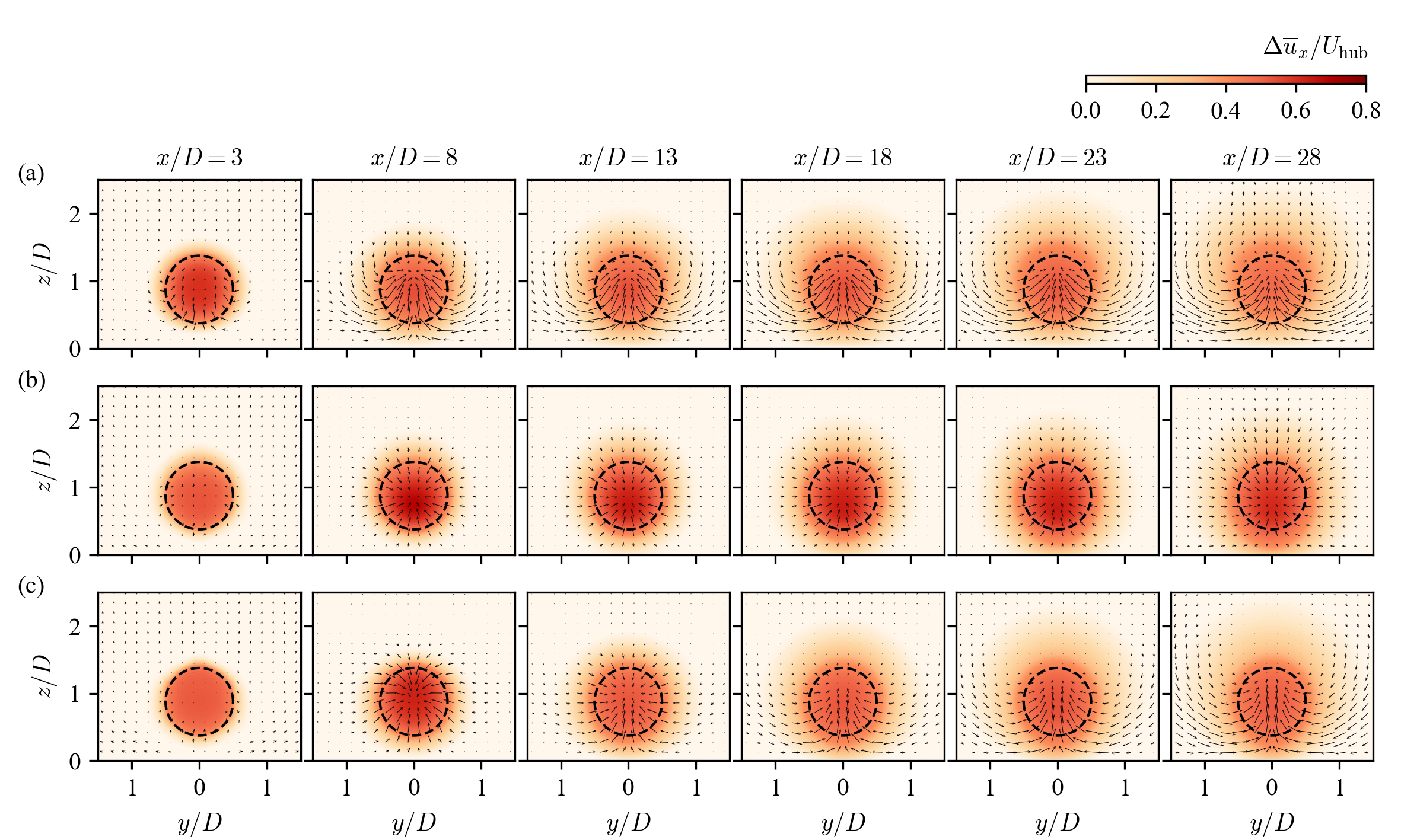}
	\caption{Contours of normalized streamwise velocity deficit and arrows of in-plane velocity, $3D$ after each turbine in Case B obtained from (a) LES, (b) $k-\omega\text{SST}$, and (c) $k-\omega\text{SST}-\text{PDA}$.}
	\label{fig:WFB_SFplots}
\end{figure}

Figure~\ref{fig:WFB_SFplots} presents the normalized streamwise velocity deficit and in-plane motion vectors in $yz$-planes, at a location of $3D$ after each turbine row in Case B. The non-linear part of the RST incorporated in $k-\omega\text{SST}-\text{PDA}$ successfully generates the secondary flow in the wake, where the linear eddy-viscosity baseline model is unable to predict such structures. Aligned with capturing this physics, we see a consistent elevation of the wake center, similar to LES. In addition to this enhancement, the $k-\omega\text{SST}-\text{PDA}$ gives closer-to-LES values for the normalized streamwise velocity deficit, especially after the third row, thanks to the corrections made to the $\omega$-transport equation. 

\begin{figure}[!ht]
	\centering
        \includegraphics[width=0.833\textwidth]{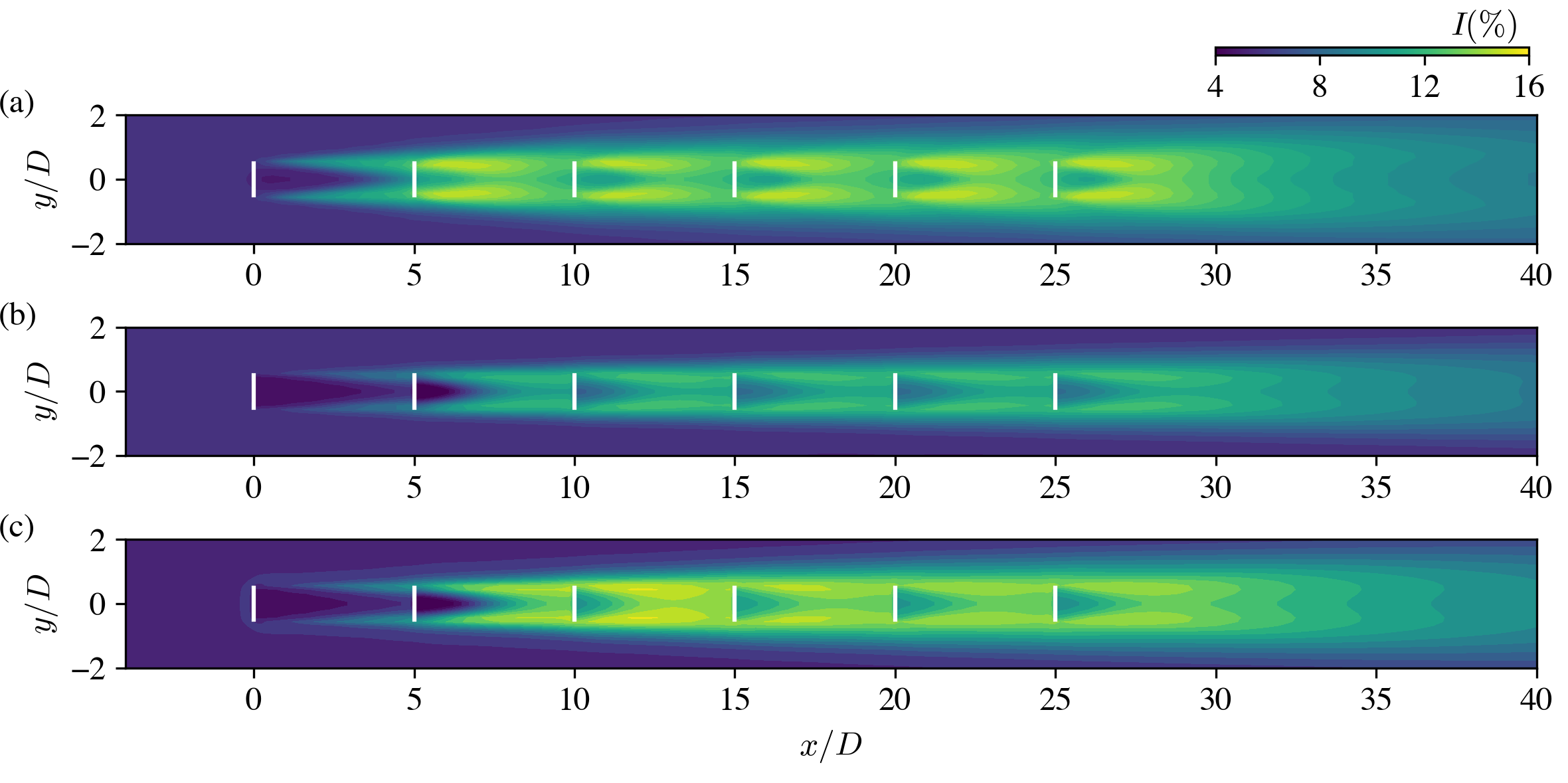}
	\caption{Contours of turbulence intensity at hub height in Case B obtained from (a) LES, (b) $k-\omega\text{SST}$, and (c) $k-\omega\text{SST}-\text{PDA}$.}
	\label{fig:WFB_Intensity}
\end{figure}

Figure~\ref{fig:WFB_Intensity} depicts the distribution of the turbulence intensity at the hub height obtained from LES, baseline model, and $k-\omega\text{SST}-\text{PDA}$ when applied to Case B. As expected, we generally observe higher values compared to Case A owing to the dense placement of turbines. Comparing Figure~\ref{fig:WFB_Intensity}(a) and (c) shows the success of the PDA approach in terms of predicting both the double-peak structures and also the magnitude of turbulence intensity in the wake flow. 

\begin{figure}[!ht]
	\centering
        \includegraphics[width=\textwidth]{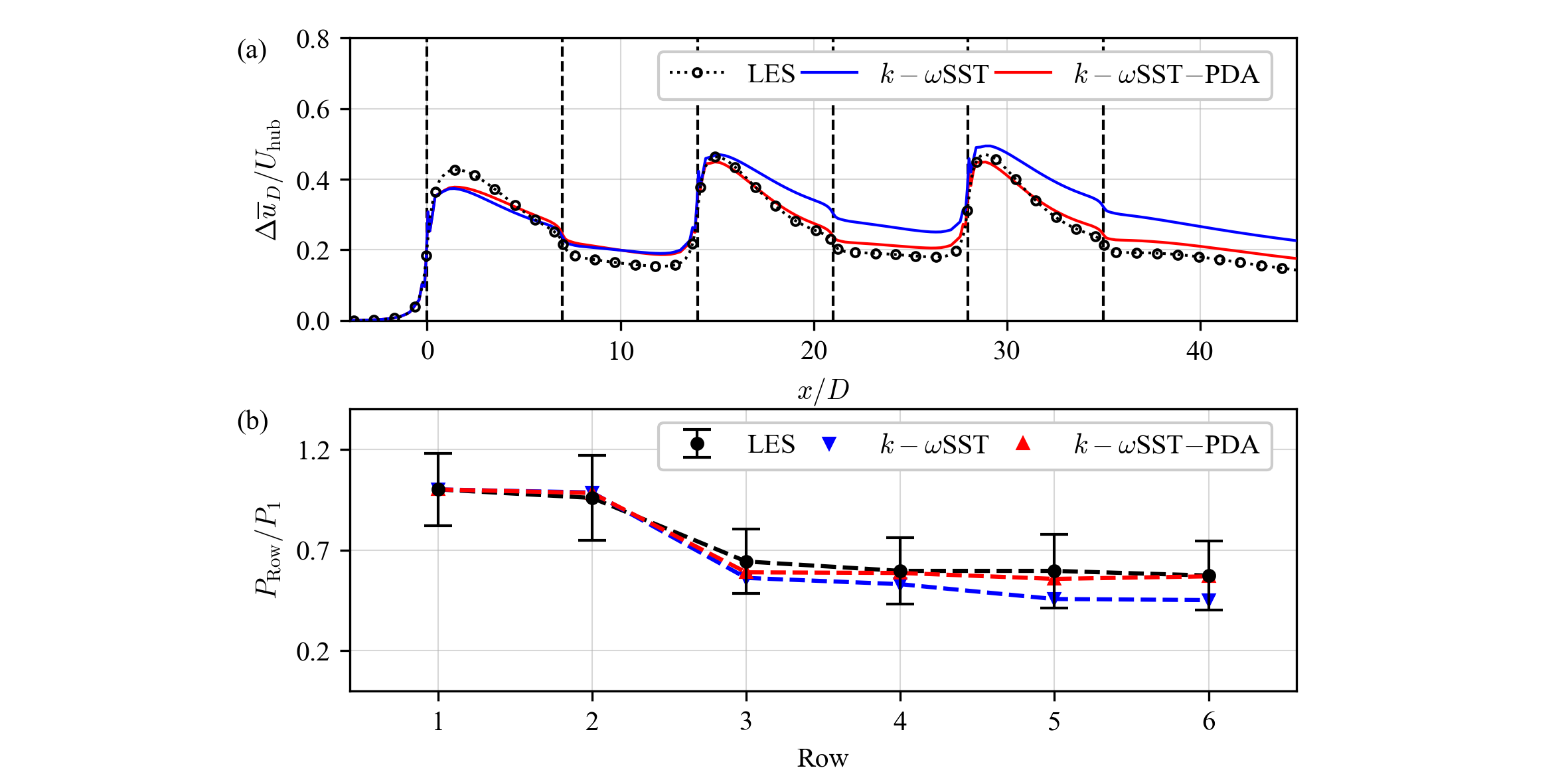}
	\caption{(a) Normalized velocity deficit,  averaged across the rotor area of odd turbine rows, and \blue (b) the normalized power for each turbine row, both given for Case C.\black}
	\label{fig:WFC_DiskQuantities}
\end{figure}

For Case C, Figure~\ref{fig:WFC_DiskQuantities}(a) presents the velocity deficit, averaged across the rotor area of the odd turbine rows. For turbine row 3 onward, the baseline model is overpredicting the velocity deficit and, consequently, power production (given in Figure~\ref{fig:WFC_DiskQuantities}(b)) is underestimated compared to the LES results. On the contrary, $k-\omega\text{SST}-\text{PDA}$ successfully improves the prediction of the velocity deficit and power output towards those from LES.

\begin{figure}[!ht]
	\centering
        \includegraphics[width=\textwidth]{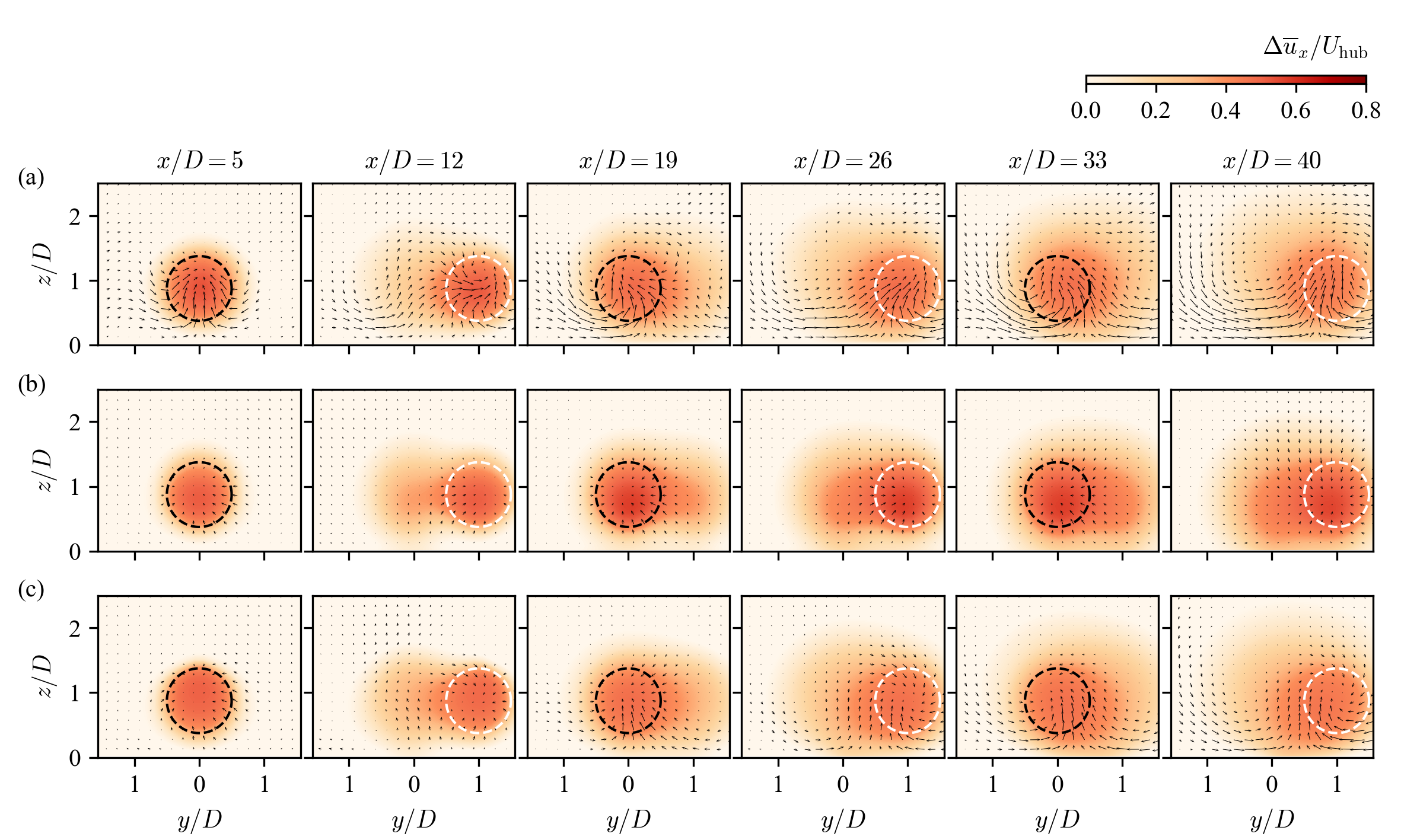}
	\caption{Contours of normalized streamwise velocity deficit and arrows of in-plane velocity, $5D$ after each turbine in Case C obtained from (a) LES, (b) $k-\omega\text{SST}$, and (c) $k-\omega\text{SST}-\text{PDA}$. Here, the black dashed circles indicate the rotors on odd rows whereas the white dashed circles indicate the shifted rotors on the even rows.}
	\label{fig:WFC_SFplots}
\end{figure}

To evaluate the success of the PDA approach in predicting the emergence of secondary flows in Case C, we illustrate the in-plane motions and normalized streamwise velocity deficit, at a distance of $5D$ after each turbine row in Figure~\ref{fig:WFC_SFplots}. Focusing on LES predictions, the secondary flow patterns are distorted toward the center of the wake after each turbine, and $k-\omega\text{SST}-\text{PDA}$, in contrast to the baseline model, can successfully improve the prediction of such features similar to LES.

\begin{figure}[!ht]
	\centering
        \includegraphics[width=0.833\textwidth]{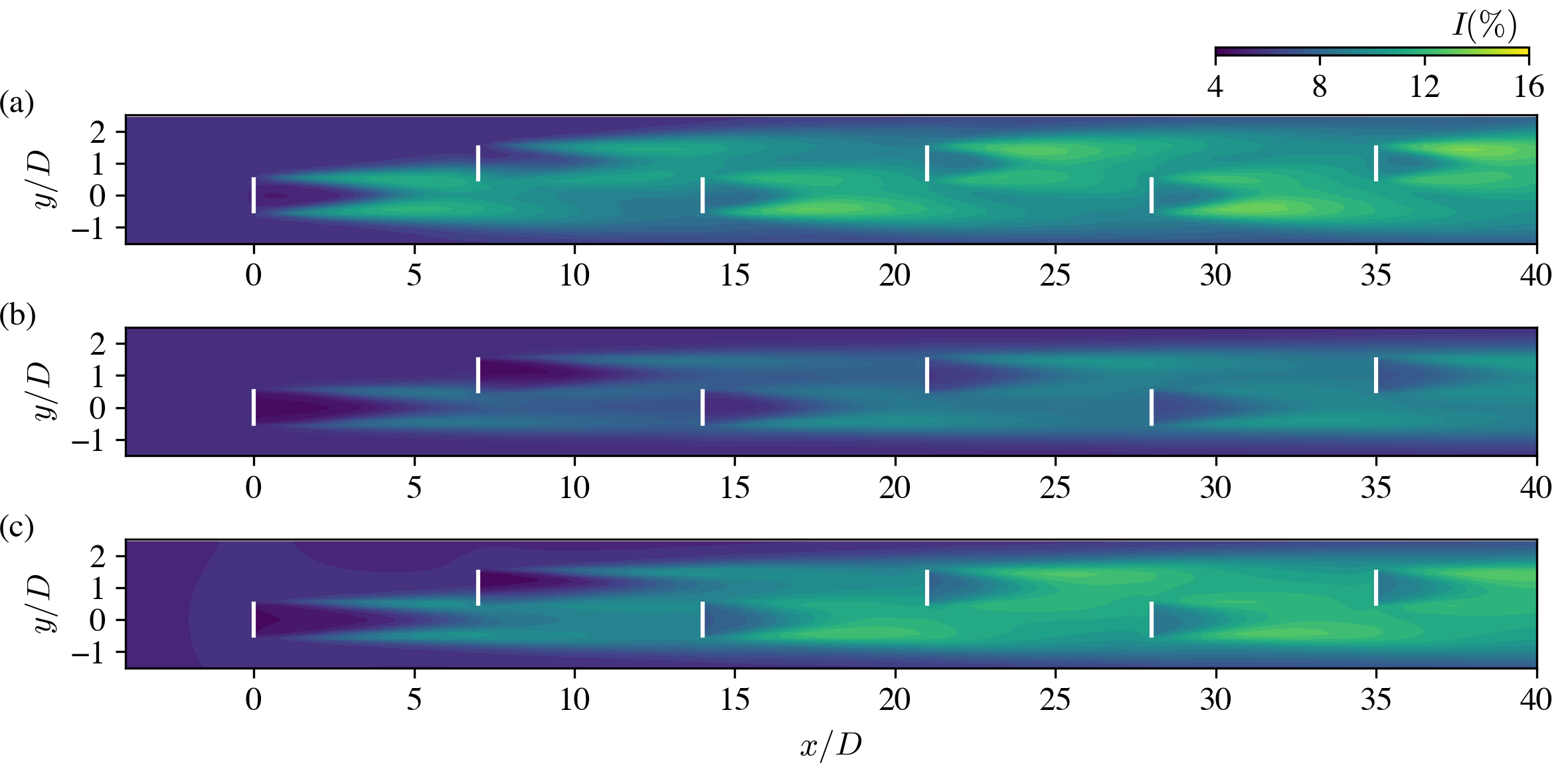}
	\caption{Contours of turbulence intensity at hub height in Case C obtained from (a) LES, (b) $k-\omega\text{SST}$, and (c) $k-\omega\text{SST}-\text{PDA}$.}
	\label{fig:WFC_Intensity}
\end{figure}

To measure the adequacy of the  $k-\omega\text{SST}-\text{PDA}$ in predicting the turbulence intensity in a case with partial wakes, in Figure~\ref{fig:WFC_Intensity}, we compare the results obtained from the three methods in a $xy$-plane at hub height. Taking the LES results as the reference reveals a diminished added turbulence within the wake in Case C, compared to Cases A and B. Notably, the figure illustrates the underestimation of turbulence intensity by the baseline model. On the other hand, the optimized $k-\omega\text{SST}-\text{PDA}$ can successfully improve the prediction of turbulence intensity, in full consistency with the two cases studied earlier.

\blue
\subsection{Validation using wind-tunnel data} 
While our model has demonstrated promising results across several validation cases, to push the boundaries of its predictive power and to broaden the scope of our analysis, we now turn to the wind-tunnel experiment conducted by Chamorro and Port{\'e}-Agel \cite{Chamorro2011}, which was carried out under neutral stratification conditions. In their setup, a $10 \times 3$ array of scaled-down wind turbines was used, with a hub height of $0.83D$, streamwise spacing of $5D$, and spanwise spacing of $4D$, where $D = 0.15 \text{ m}$. Table~\ref{tab:Ct} presents the $C_T$ and $C_T^\prime$ values for different turbine rows, distinct from those in Cases A, B, and C. The roughness height and boundary-layer depth in these experiments were reported as $19.98 \times 10^{-5} D$ and $4.5D$, respectively \cite{Chamorro2011}. In addition to the measurement data, we incorporate results from LESs conducted by Stevens \textit{et al.} \cite{Stevens2018} to serve as an additional benchmark for comparison. We simulate the entire wind farm using our RANS framework in conjunction with both $k-\omega\text{SST}$ and $k-\omega\text{SST}-\text{PDA}$ models. For brevity, details regarding the computational domain and boundary conditions are omitted here but can be found in Refs. \cite{zehtabiyanrezaie2023extended, zehtab2024secondary}.

\begin{figure}[!ht]
	\centering
        \includegraphics[width=\textwidth]{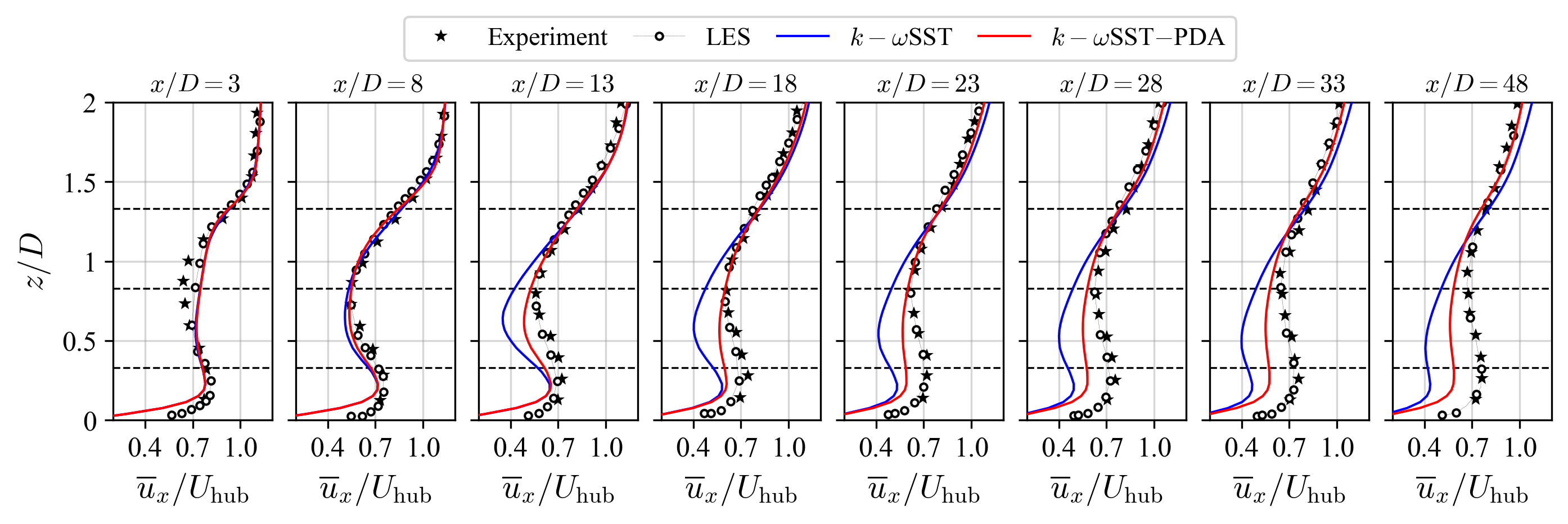}
         \caption{\blue Normalized velocity profiles measured vertically at a downstream distance of $3D$ from rows 1-7 and 10 in the $10 \times 3$ wind-turbine array at $y = 0$. The profiles are predicted by $k-\omega\text{SST}$ and $k-\omega\text{SST}-\text{PDA}$, and are compared with wind-tunnel measurement data and LES results. The wind tunnel data (black stars) are derived from the experiments by Chamorro and Port{\'e}-Agel \cite{Chamorro2011}. The measurement data and LES results are sourced from Ref. \cite{Stevens2018}. Horizontal black dashed lines mark the bottom-tip, hub, and top-tip heights.\black}

	\label{fig:windTunnelComp}
\end{figure}

Figure~\ref{fig:windTunnelComp}  shows the vertical profiles of normalized streamwise velocity obtained from the RANS models, compared against wind-tunnel measurements \cite{Chamorro2011} and LES data \cite{Stevens2018}. These profiles are taken at a distance of $3D$ downstream of turbine rows 1 through 7 and row 10. One can see that the adjustments to the $\omega$-transport equation, along with the incorporation of secondary flows in the wake, result in a consistently better agreement with high-fidelity data across the entire vertical extent, from the bottom to the tip.\black

\begin{table}[h]
    \centering
    \caption{\blue Values of $C_T$ and $C_T^\prime$ for various turbine rows utilized in the RANS simulation of the $10 \times 3$ array, as derived from Ref. \cite{Stevens2018}.\black}
    \label{tab:Ct}
    \begin{tabular}{ccc|ccc}
    \hline
    Row & $C_{T}$ & $C_{T}^\prime$ & Row & $C_{T}$ & $C_{T}^\prime$ \\
    \hline
    1  & 0.5091 & 0.7041 & 6  & 0.6202 & 0.9496 \\
    2  & 0.5601 & 0.8099 & 7 & 0.6109 & 0.9269 \\
    3  & 0.6406 & 1.0015 & 8 & 0.5898 & 0.8768 \\
    4  & 0.6116 & 0.9286 & 9 & 0.5926  & 0.8831 \\
    5  & 0.5912 & 0.8799 & 10 & 0.5955 & 0.8899 \\
    \hline
    \end{tabular}
\end{table}

\section{Conclusions} \label{Sec:Conclusions}
This study has been dedicated to presenting a progressive data-augmented model for enhanced Reynolds-averaged simulations of wind farms. \blue It is well established that the commonly used empirical models struggle to precisely predict Reynolds-stress tensors in the wake \black and, at the same time, cannot capture intricate flow phenomena linked to turbulence anisotropy such as secondary flows of Prandtl’s second kind behind the turbines. To address these challenges, firstly, we built a baseline model by implementing the impact of turbine forces in the TKE equation of the well-known $k-\omega\text{SST}$ model. We then leveraged LES data, progressively augmenting the capability of the baseline model to satisfactorily predict both the eddy viscosity and turbulence anisotropy in the wake flow.

We considered three wind-farm cases with different layouts in this study, where one case was used for the optimization and the other two cases were used for the validation of the developed model. The optimal coefficients within the core of the data-augmented progressive model were found by a grid-search optimization technique. The coefficients were associated with two corrective terms. One of these terms refined the $\omega$-transport equation to improve eddy-viscosity predictions, while the other, acting as a nonlinear term within RST, enabled the emergence of secondary flows.

We used data from LESs conducted in-house as our high-fidelity reference for the evaluation of the new model's performance. Table~\ref{tab:Table1} summarizes the results obtained by the new model compared to the baseline model. In the optimization case with 6 inline turbine rows with a spacing of $7D$, the optimized model showed a successful improvement in the prediction of streamwise velocity deficit which led to a better prediction of power output. \blue The new progressive model also reduced the error values of the in-plane velocity components from $62\%-73\%$ to $40\%-53\%$ in all cases which quantitatively confirms the improvement in prediction of the secondary flows. \black Also, a notable improvement was observed in the prediction of turbulence intensity by using the new progressive model compared to the original $k-\omega\text{SST}$. 
\blue 
It should be mentioned that the new corrections include only a few extra algebraic calculations, which did not significantly impact the computational cost of the RANS simulation compared to the original $k-\omega\text{SST}$ model.
\black

\begin{table}
\centering
\caption{A summary of the performance of the  baseline model ($k-\omega\text{SST}$) and $k-\omega\text{SST}-\text{PDA}$ when applied to wind-farm cases considered in this study.}
\label{tab:Table1}
\setlength{\tabcolsep}{6pt}
\begin{tabular}{ccccccc}
\hline
& \multicolumn{2}{c}{Case A (Optimization)} & \multicolumn{2}{c}{Case B (Validation)} & \multicolumn{2}{c}{Case C (Validation)}\\
Averaged error($\%$) on   & baseline & baseline + PDA & baseline & baseline + PDA & baseline & baseline + PDA \\ \hline
$\Delta \overline{u}_x$        & 40.65 & 20.48  & 40.09 & 22.08  & 32.06 & 17.94 \\  
$\overline{u}_y$               & 67.90 & 40.25  & 62.58 & 39.87  & 65.41 & 52.50 \\   
$\overline{u}_z$               & 73.66 & 45.37  & 68.25 & 53.02  & 70.98 & 47.37 \\    
$I$                 & 26.87 & 10.12  & 24.27 & 10.29  & 26.77 & 9.58 \\    
Power output    & 21.77 & 6.74   & 23.73 & 8.92   & 9.97  & 3.02 \\ \hline   

\end{tabular}
\end{table}

For the validation of the $k-\omega\text{SST}-\text{PDA}$ model, we then evaluated its performance on two unseen cases including Cases B and C. In Case B, the distancing between each two rows of turbines was reduced to $5D$, where turbines were under a stronger wake effect. In Case C, the streamwise distancing remained $7D$ but the even rows of turbines were shifted in the spanwise direction; therefore, downstream turbines were experiencing a partial wake effect. 
\blue A closer look at the tendency of reduction in error values in Table~\ref{tab:Table1} shows that a similar improvement has occurred both for seen Case A and unseen Cases B and C; for example, the error in prediction of the turbulent intensity in all three cases reduced from $24\%-27\%$ to $10\%$. This similar pattern in improvements shows the high level of generalizability of the progressive RANS model in the simulation of wind-farm wakes. \black

\blue Expanding our investigation to a $10 \times 3$ array of wind turbines, we employed wind-tunnel experiments \cite{Chamorro2011} and LES data \cite{Stevens2018} to evaluate the performance of our model beyond the initial validation scenarios. The findings revealed a notable enhancement in prediction accuracy relative to reference data, underscoring the efficacy of our approach and robustness of the methodology\black.

\blue 
This study demonstrated the successful application of a progressive data-driven turbulence model for wind-farm flow simulations, using the $k-\omega\text{SST}$ model as the baseline. The coefficients $C_{P\omega}$ and $C_{\text{SF}}$ were specifically optimized for this baseline model. Future research could focus on fine-tuning these coefficients for other baseline models to enhance the model’s versatility and performance across a broader range of scenarios.
\black
\black

\section*{CRediT authorship contribution statement}
\textbf{Ali Amarloo}: Conceptualization, Methodology, Software, Formal analysis, Investigation, Data curation, Visualization, Writing – original draft. \textbf{Navid Zehtabiyan-Rezaie}: Conceptualization, Methodology, Software, Formal analysis, Investigation, Data curation, Visualization, Writing – original draft.  \textbf{Mahdi Abkar}: Conceptualization, Funding acquisition, Methodology, Project administration, Resources, Supervision, Writing – review \& editing. 
\section*{Conflict of interest}
The authors have no conflicts to disclose.
\section*{Data availability statement}
The data that support the findings of this study are available from the corresponding author upon reasonable request.
\section*{Acknowledgments}
The authors acknowledge the financial support from the Independent Research Fund Denmark (DFF) under Grant No. 0217-00038B. The computational resources used in this study were provided by the Danish e-Infrastructure Cooperation (DeiC) National HPC under Project No. DeiC-AU-N2-2023009 and DeiC-AU-N5-2023004.
\appendix
\section{Grid-convergence study}\label{app:App}
\begin{figure}[!ht]
	\centering
        \includegraphics[width=\textwidth]{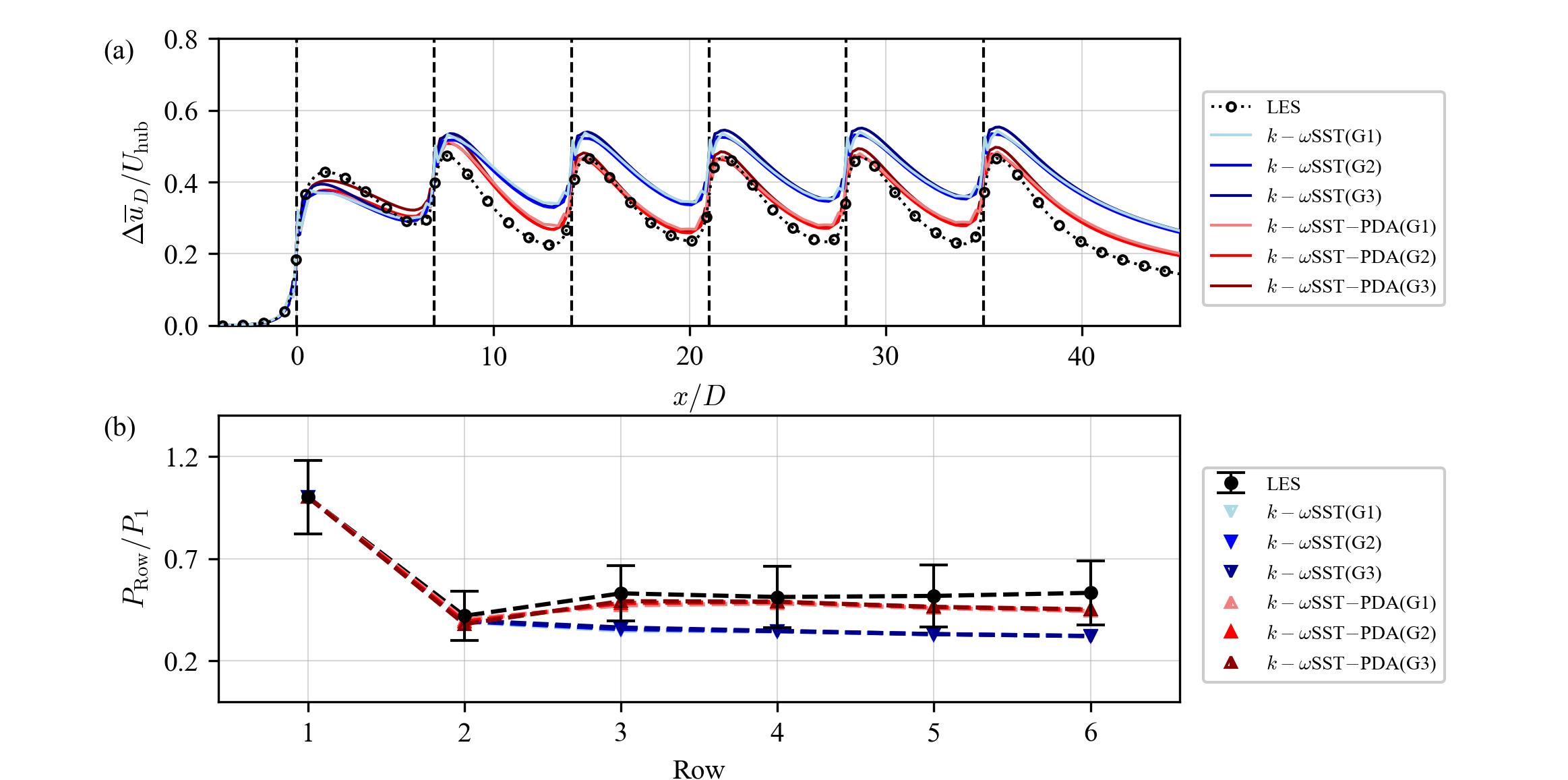}
	\caption{\blue (a) Normalized velocity deficit,  averaged across the rotor area of odd turbine rows, and (b) the normalized power for each turbine row, both given for Case A for different grid resolutions.\black}
    \label{fig:GridConv}
\end{figure}

\blue To assess the sensitivity of our results to grid resolution, we compare three different grid sizes: $165 \times 28 \times 42$ (G1), $234 \times 40 \times 58$ (G2), and $306 \times 57 \times 83$ (G3). Figure~\ref{fig:GridConv}(a) illustrates the variation in rotor-averaged normalized velocity deficit with downstream distance for these grid configurations in Case A. As Cases B and C exhibit similar trends, their results are omitted for brevity. Our analysis reveals that all grid resolutions yield comparable results for the velocity deficit. Likewise, Figure~\ref{fig:GridConv}(b) indicates that normalized power production remains largely unaffected by changes in grid resolution. Therefore, we adopt the medium-resolution grid, G2, for the analysis throughout this paper.
 \black

\bibliographystyle{elsarticle-num} 

\end{document}